\documentclass[usenatbib]{mn2e}
\usepackage{graphicx}
\usepackage{ifthen}
\usepackage{amsmath}
\usepackage{url}
\usepackage{rotating}
\usepackage{xcolor}

\def\ltsima{$\; \buildrel < \over \sim \;$}
\def\lta{\lower.5ex\hbox{\ltsima}}
\def\gtsima{$\; \buildrel > \over \sim \;$}
\def\simgt{\lower.5ex\hbox{\gtsima}}
\def\kms{{\rm\,km \; s^{-1}}}

\def\kpc{{\rm\,kpc}}

\def\msun{{\rm\,M_\odot}}
\def\lsun{{\rm\,L_\odot}}

\def\AA{$\; \buildrel \circ \over {\rm A}$}

\def\s{\ifmmode \widetilde \else \~\fi}
\def\={\overline}

\def\spose#1{\hbox to 0pt{#1\hss}}

\def\lta{\mathrel{\spose{\lower 3pt\hbox{$\mathchar"218$}}
     \raise 2.0pt\hbox{$\mathchar"13C$}}}
\def\gta{\mathrel{\spose{\lower 3pt\hbox{$\mathchar"218$}}
     \raise 2.0pt\hbox{$\mathchar"13E$}}}
\def\Dt{\spose{\raise 1.5ex\hbox{\hskip3pt$\mathchar"201$}}}    % upper case
\def\dt{\spose{\raise 1.0ex\hbox{\hskip2pt$\mathchar"201$}}}    % lower case

\def\dotsfill{\leaders\hbox to 1em{\hss.\hss}\hfill}

\def\FeH{{\rm[Fe/H]}}

\loadboldmathitalic 
\title{Detailed study of the Milky Way globular cluster Laevens~3.}

\author[N. Longeard et al.] {Nicolas Longeard$^{1}$, Nicolas Martin$^{1,2}$, Rodrigo A. Ibata$^{1}$, Michelle L. M. Collins$^{3}$
\newauthor Benjamin P. M. Laevens$^{4}$, Eric Bell$^{5}$, Dougal Mackey$^{6}$ \\
$^{1}$ Universit\'e de Strasbourg, CNRS, Observatoire astronomique de Strasbourg, UMR 7550, F-67000 Strasbourg, France\\
$^{2}$ Max-Planck-Institut f\"ur Astronomy, K\"onigstuhl 17, D-69117, Heidelberg, Germany\\
$^{3}$ Department of Physics, University of Surrey, Guildford, GU2 7XH, Surrey, UK\\
$^{4}$ Institute of Astrophysics, Pontificia Universidad Cat\'olica de Chile, Av. Vicuña Mackenna 4860, 7820436 Macul, Santiago, Chile\\
$^{5}$ Department of Astronomy, University of Michigan, 500 Church St., Ann Arbor, MI 48109, USA\\
$^{6}$ Research School of Astronomy and Astrophysics, Australian National University, Canberra, ACT 2611, Australia\\
}

\date{\today}
\begin{document} 
\maketitle 

\begin{abstract} 

We present a photometric and spectroscopic study of the Milky Way satellite Laevens~3. Using MegaCam/CFHT $g$ and $i$ photometry and Keck II/DEIMOS multi-object spectroscopy, we refine the structural and stellar properties of the system. The Laevens~3 colour-magnitude diagram shows that it is quite metal-poor, old ($13.0 \pm 1.0$ Gyr), and at a distance of $61.4 \pm 1.0$ kpc, partly based on two RR Lyrae stars. The system is faint ($M_V = -2.8^{+0.2}_{-0.3}$ mag) and compact ($r_h = 11.4 \pm 1.0 $ pc). From the spectroscopy, we constrain the systemic metallicity ($\FeH_\mathrm{spectro} = -1.8 \pm 0.1$ dex) but the metallicity and velocity dispersions are both unresolved. Using Gaia DR2, we infer a mean proper motion of $(\mu_\alpha^*,\mu_\delta)=(0.51 \pm 0.28,-0.83 \pm 0.27)$ mas yr$^{-1}$, which, combined with the system's radial velocity ($\langle v_r\rangle = -70.2 \pm 0.5 \kms$), translates into a halo orbit with a pericenter and apocenter of $40.7 ^{+5.6}_{-14.7}$ and $85.6^{+17.2}_{-5.9}$ kpc, respectively. Overall, Laevens~3 shares the typical properties of the Milky Way's outer halo globular clusters. Furthermore, we find that this system shows signs of mass-segregation which strengthens our conclusion that Laevens~3 is a globular cluster. 
\end{abstract}


\section{Introduction}

In recent years, the faint regime of Milky Way (MW) satellites has been explored under the impulsion of large photometric surveys. Among those, we can cite the Sloan Digital Sky Survey \citep{york2000}, the Panoramic Survey Telescope and Rapid Response System 1 \citep{chambers16} or the Dark Energy Survey \citep{abbott05}. These surveys led to numerous discoveries of faint satellites. Several old and metal-poor faint systems have been identified as globular clusters (GCs) (\citealt{balbinot13}, \citealt{laevens14}, \citealt{kim15b}, \citealt{kim16b}), although some of them require confirmation \citep{martin16c}. Because of their old stellar populations, they can be considered as the witnesses of the formation of their host galaxy \citep{strader05} and bring insights on low-mass galaxy formation. Furthermore, the chemodynamics of those GCs can also trace some of the current properties of their host \citep{pota13}. GCs can also be useful to constrain stellar population models \citep{chantereau16}. The fact that these diffuse and small satellites survived for several billion years can also bring more information on their formation and internal processes (\citealt{baumgardt_makino03}, \citealt{renaud17}).

The GCs associated with the MW span a wide range of luminosities, metallicities and distances \citep{harris10}, but only a few have been discovered in the outer reaches of the halo ($R_\mathrm{gal} > 50 \kpc$). This specific group of clusters is in fact suspected to not have formed in-situ, but rather as companions in nearby dwarf galaxies and accreted at later times in the MW history (\citealt{mackey10}, \citealt{dotter11}).  While clusters like Pal 14 (\citealt{arp_vanderbergh60}, $d_\mathrm{gal} \sim 71 \kpc$) or AM-1 (\citealt{madore_arp79}, $d_\mathrm{gal} \sim 125 \kpc$) have been known for decades, only a handful of fainter outer halo clusters were discovered in recent photometric surveys. Laevens~1/Crater (\citealt{belokurov14}, \citealt{laevens14}) and Kim 2 \citep{kim15} fall in this category. Such faint satellites often lie in the so-called ``valley of ambiguity'' where the frontier between dwarf galaxies and old stellar clusters is not clearly defined \citep{gilmore07}. Laevens~1 is a great illustration of that, as its very nature was disputed at the time of its discovery. Indeed, while \citet{laevens14} identified the system as a cluster, \citet{belokurov14} proposed that the satellite could have been a tidally disrupted dwarf galaxy. This example only accentuates the hardship of studying these faint, distant stellar systems. In such an extreme regime, the combination of photometric, chemical, and kinematics data is needed to both classify and understand those systems. 

Laevens 3 (Lae~3) is a system first discovered in the Pan-STARRS~1 \citep[PS1]{chambers16} data by \citet{laevens15}. At the time, it was found to be compact ($r_h = 7 \pm 2 $ pc) and the existence of an RR Lyrae star in this region, probably belonging to the system, allowed to constrain the distance to the system ($64 \pm 3 $ kpc). Using this distance, \citet{laevens15} found that the main sequence of Lae~3 was compatible with a stellar population of 8 Gyr, and a metallicity of $\FeH = -1.9$. From these properties, the authors concluded that the system is a faint Milky Way globular cluster.

In this work, we undertake a careful refinement of the properties of the satellite through deep broadband photometry with CFHT/MegaCam, as well as the first spectroscopic follow-up of the system using Keck/DEIMOS \citep{faber03}. Section 2 discusses the technical aspects of our observations. Section 3 details the photometric analysis that derives the structural and colour-magnitude diagram (CMD) properties of the satellite. In section 4, we present the dynamics of Lae~3 using multi-object spectroscopy, while section 5 details the orbital properties of the satellite obtained with the Gaia Data Release 2 data. Finally, the nature and main properties of Lae~3 are discussed in section 6. 

\section{Observations}

\subsection{Photometry}

\begin{figure*}
\begin{center}
\centerline{\includegraphics[width=\hsize]{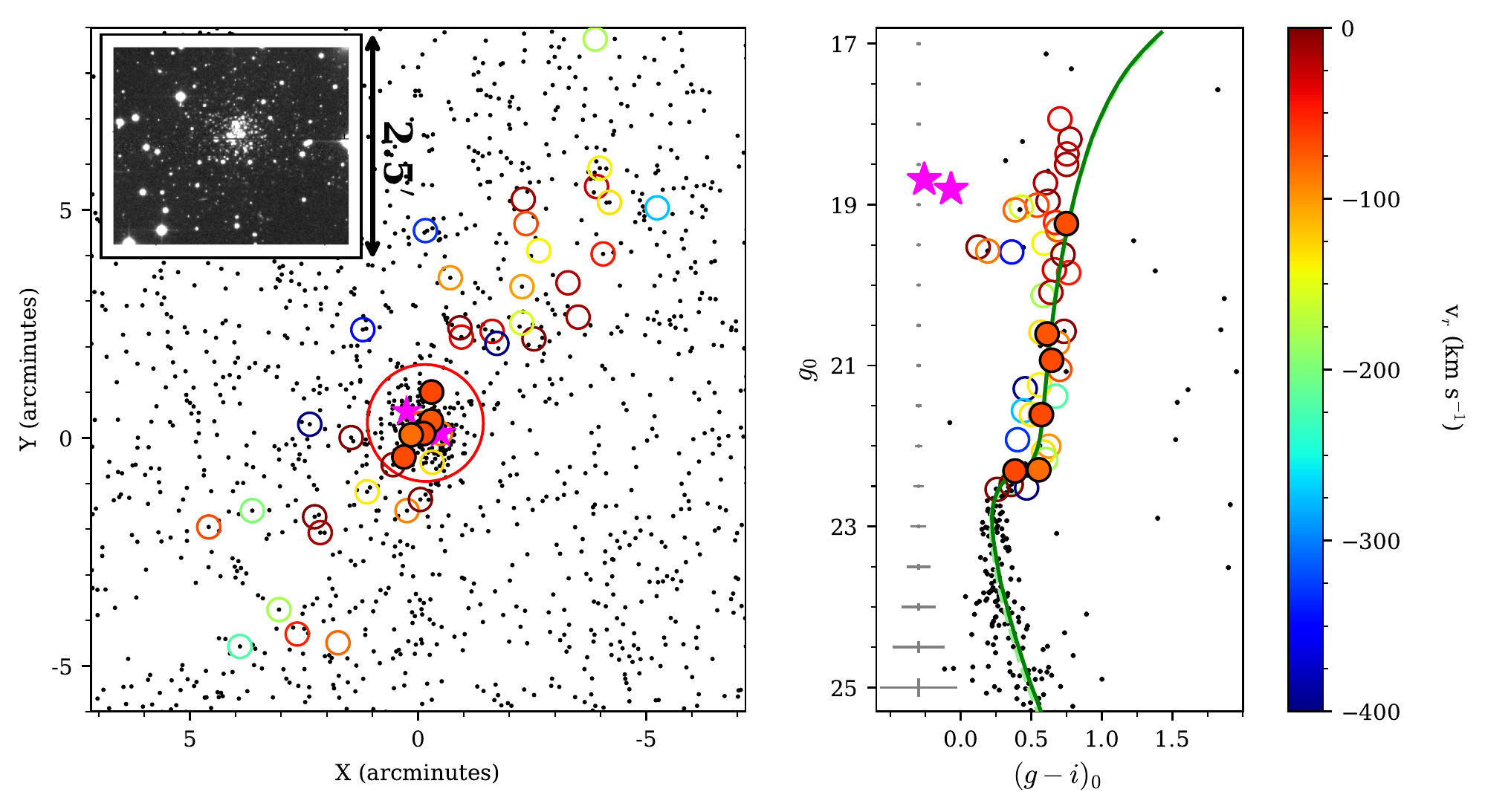}}
\caption{{\textit{Left panel: }}Spatial distribution of the Lae~3-like stellar population in the field of view. The CFHT image of the 2.5$'$x2.5$'$ region around Lae~3 in the $i$ band is shown in the upper-left corner. The red circle represents the two half-light radii ($r_h \sim 0.64'$) region of Lae~3. The two RR Lyrae identified in the system are shown as magenta stars. The spectroscopic dataset is represented by circles, colour-coded according to their heliocentric velocities. Filled circles stand for stars identified as Lae~3 members. \textit{Right panel: }CMD within two half-light radii of Lae~3. The best fitting isochrone derived in section 3.1 is represented as a solid green line, while the stellar population inferred without any distance or metallicity priors is represented by the light green dashed line. Photometric uncertainties are reported as grey error bars on the left side of the plot.}
\label{one}
\end{center}
\end{figure*}

\begin{figure*}
\begin{center}
\centerline{\includegraphics[width=\hsize]{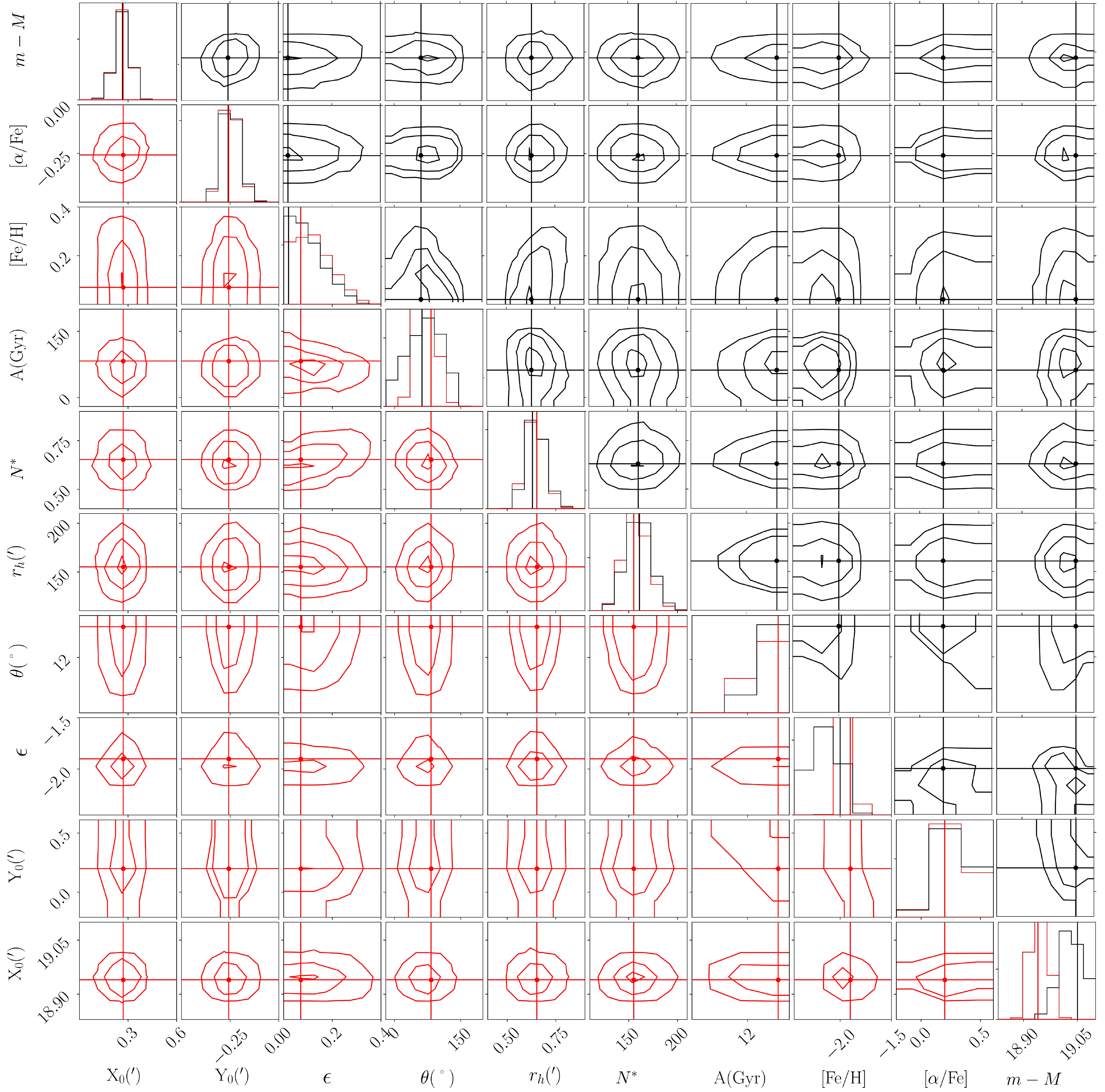}}
\caption{One- and two-dimensional posterior PDFs of the structural and CMD parameters of Lae~3, inferred using the method described in section 3.1. Contours correspond to the usual 1, 2 and 3$\sigma$ confidence intervals in the case of a two-dimensional Gaussian. The red solid lines correspond to the analysis using the both the distance and metallicity priors described in section 3, while the black lines represent the case without any prior applied. Black and red dots correspond to the favoured model  in each case.}
\label{cornerplots}
\end{center}
\end{figure*}

\begin{figure}
\begin{center}
\centerline{\includegraphics[width=\hsize]{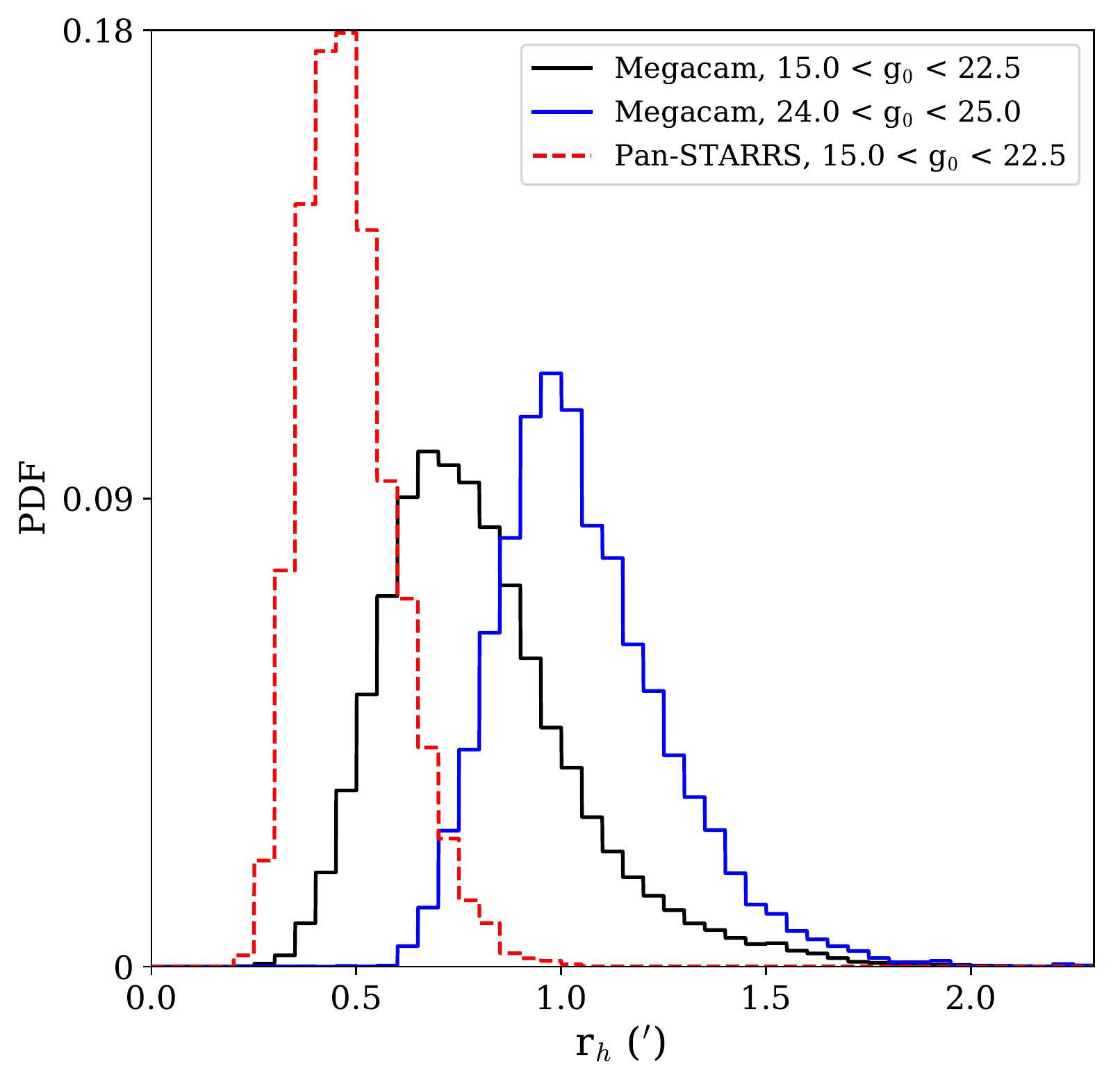}}
\caption{One-dimensional PDFs of the half-light radius of Lae~3 in three cases: using stars with $ 15.0 < g_0 < 22.5$ in MegaCam (solid black line), stars with $ 24.0 < g_0 < 25.0$ (solid blue line) in MegaCam, and the PS1 catalog (dashed red line). The magnitude ranges in the first two cases were chosen so that the inferred numbers of Lae~3 stars are similar. Lae~3 comes out as larger when considering lower mass stars than when the analysis is performed on a more massive sample, hinting at a mass-segregation process. The size of the satellite inferred by L15 is retrieved when using their data, indicating that this effect is not caused by a problem in our approach or a statistical fluke. }
\label{pdfs_rh}
\end{center}
\end{figure}

The photometry used in this work consists of multi-exposures MegaCam broadband $g$- and $i$-band images. The exposure times are of $3 \times 480$ s for $g$ and $3 \times 540$ s for $i$. The observations were conducted in service mode by the CFHT crew during the night of July 18th, 2015 under excellent seeing conditions ($\sim 0.3 ''$), and the data reduced following the procedure detailed in \citet[L18]{longeard18}. We use the Cambridge Astronomical Survey Unit (CASU, \citealt{irwin01}) pipeline flags to perform the star/galaxy separation. CASU also indicates all saturated sources. The calibration of the MegaCam  photometry \citep{boulade03} is performed onto the PS1 photometric system similarly to L18. We first cross-identified all unsaturated point sources between PS1 and MegaCam. Only stars with photometric uncertainties below 0.05 in both catalogs are then considered for the calibration. We assume that the transformation between the PS1 and MegaCam photometry can be reliably modelled by a second-order polynomial, with a 3-$\sigma$ clipping procedure.

All stars saturated in the MegaCam photometry are directly imported from the PS1 catalog, for a total of 51,759 stars. Finally, the catalog is dereddened using the 2D dust map from \citet{schlegel98} to determine the line-of-sight extinction and \citet{schlafly11} for the extinction  coefficients.

\subsection{Spectroscopy}

The spectroscopic run for Lae~3 was performed on the night of Sept 7, 2015 (Julian date of 2457272.5) using Keck II/DEIMOS. The targets were selected based on their distance to Lae~3 and their location on the colour-magnitude diagram, using the PS1 photometry presented in \citet{laevens15}. A total of 51 stars were observed using the OG550 filter and the 1200 lines mm$^{-1}$ grating. The typical central wavelength resolution is $R \sim 8500$, covering the spectral range from $6500$ to $9000$ \AA. The spectra were then processed using the IRAF SIMULATOR package from the Keck Observatories and the pipeline detailed in \citet{ibata11}. Stars with a signal-to-noise ratio below 3 as well as the ones with radial velocity uncertainties above $15\kms$ are discarded from the spectroscopic catalog. The resulting catalog consists of 44 stars for which the spatial and CMD distributions are shown in Figure \ref{one}. Finally, the instrumental systematic velocity uncertainty is chosen to be the same as in \citet{longeard19}, with $\delta_{\mathrm{thr}} = 1.8^{+0.3}_{-0.2} \kms$.

\section{Broadband photometry analysis}

The region including Lae~3 is shown in the left panel of Figure \ref{one}, with the stars observed spectroscopically colour-coded by their velocities. The central region of the system is densely populated. The colour-magnitude diagram within two half-light radii of Lae~3 is shown in the right panel of Figure \ref{one}. The great depth of the MegaCam photometry allows us to probe the system two magnitudes below the Main Sequence Turn-Off (MSTO) and clearly reveals the main sequence of Lae~3. Our spectroscopic sample extends all the way down to the sub-giant branch, and suggests that Lae~3 possesses at least a few Red Giant Branch (RGB) stars. Four RR Lyrae stars are located in the vicinity of the satellite according to the catalog of \citet{sesar17}. Among those, only stars with a RRab classification score greater than 90 per cent are selected, as the distance modulus measurement of RRc stars can be biased. Two stars pass this criterion and have a $m-M$ of $18.87 \pm 0.06$ and $18.89 \pm 0.06$ mag respectively. By doing the mean of these two distance modulii, we obtain a distance modulus estimate of $18.88 \pm 0.04$ mag for Lae~3 ($59.7^{+0.2}_{-1.0}$ kpc in physical distance).

\subsection{Structural and CMD fitting}
We aim to derive the structural and stellar population properties of Lae~3. As such, we rely on the technique presented in \citet{martin16} and L18. The stellar population parameters that we aim to infer are the age $A$, metallicity $\FeH_\mathrm{CMD}$, the $\alpha$ abundance ratio $[\alpha/Fe]$, and the distance modulus $m - M$. The structural properties that are determined are the spatial offsets of the centroid from the literature values ($\alpha = +316.72635^{\circ}$, $\delta = +14.98000^{\circ}$) $X_0$ and $Y_0$, the ellipticity $\epsilon$\footnote{The ellipticity is defined as $\epsilon = 1 - \frac{a}{b}$, with $a$ and $b$ the major and minor axes of the ellipse respectively.}, the half-light radius $r_h$, the position angle of the major axis $\theta$, and the number of stars $N*$ of the system within the dataset.

\begin{figure}
\begin{center}
\centerline{\includegraphics[width=\hsize]{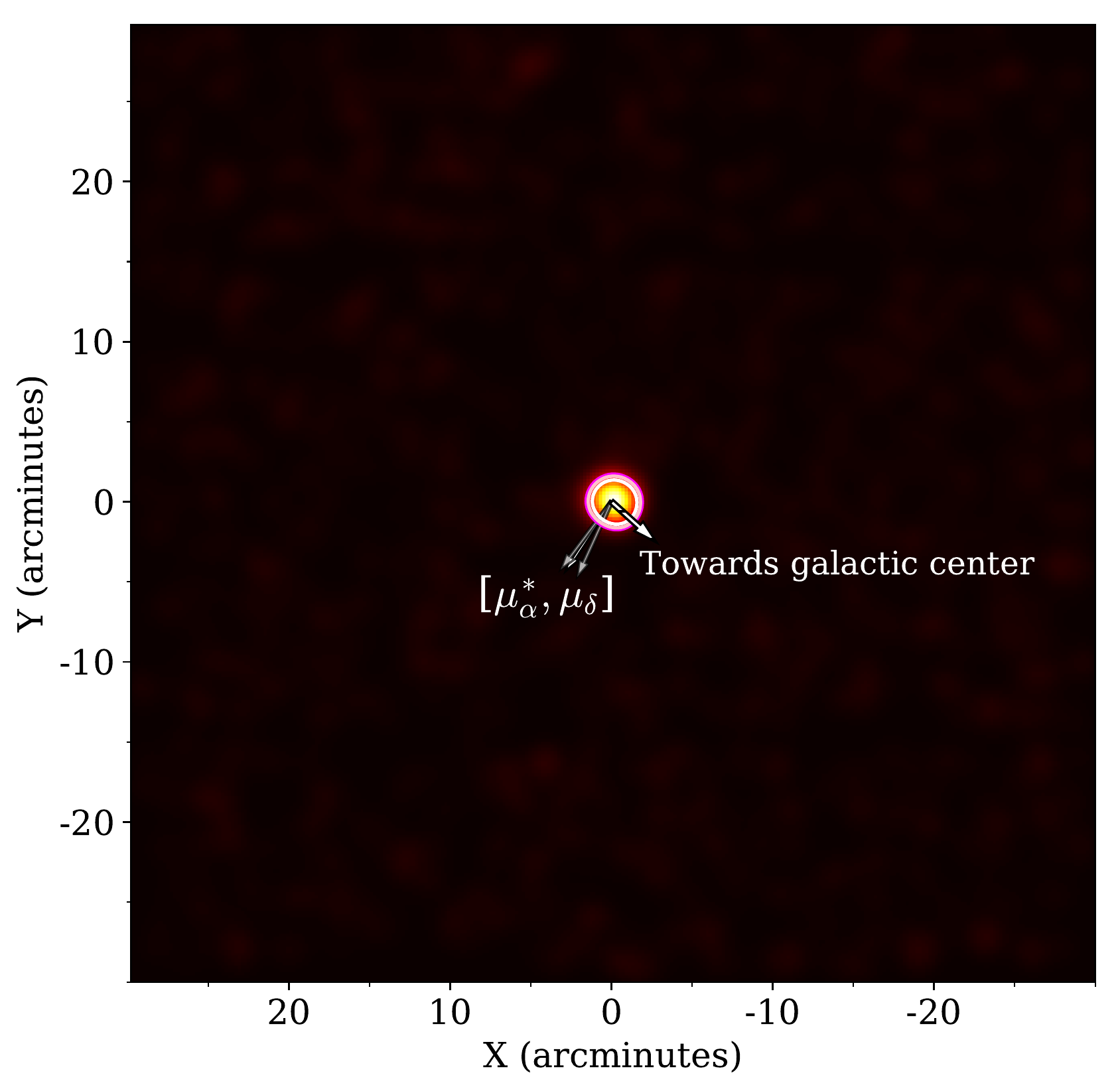}}
\caption{Density plot for all stars with P$_\mathrm{mem} > 0.01$ over the field of view. The magenta, pink and white lines outline the regions with a density higher than 68, 95 and 99 per cent of the background pixel distribution. The proper motion of Lae~3 is shown with the grey arrows along with its uncertainties, while the direction towards the galactic center is indicated with a white arrow. No tidal features is detected in the vicinity of the satellite.}
\label{density_plot}
\end{center}
\end{figure}

To derive the structural parameters, the satellite is assumed to follow an exponential radial density profile, while the spatial density of the background is assumed to be constant over the field. The stellar characteristics are determined by assuming that the CMD of the satellite can be considered as the sum of two components: a unique stellar population for Lae~3, and a contamination from the foreground stars. Given the appearance of the Lae~3 sequence in Figure \ref{one}, these assumptions are reasonable as the differences between isochrones in the metal-poor regime are not significant, except in the case of important spreads in both age and metallicity. The modelling of the CMD contamination is done empirically, by selecting all stars outside 5$r_h$ of the system. The CMD of this subsample is further binned and smoothed by a gaussian kernel of 0.1 in both colour and magnitude. The Lae~3 stellar population is, on the other hand, modelled using old and metal-poor isochrones from the Darmouth library \citep{dotter08}. The Lae~3 likelihood model is built by convolving each isochrone track by the typical photometric uncertainties of the data at a given ($g_0$, $i_0$). This model is then weighted by both the luminosity function of the track considered, and the completeness of the data at a given ($g_0$, $i_0$). This method is discussed in further details in L18. 

The distance inferred using the RR Lyrae in the field can be used as a prior for our analysis. Moreover, and anticipating on section 4, the spectroscopic analysis of three bright Lae~3 member stars allows us to infer the metallicity of the satellite to be $<\FeH_\mathrm{spectro}> = -1.8 \pm 0.1$ dex. The PDF of this result can also be used as a prior. 

The structural and CMD parameters are inferred all together and the results are displayed in Table 1, while the PDFs are shown in figure \ref{cornerplots}. We find that Lae~3 is spherical, with a half-light radius of $0.64 \pm 0.05$ arcminutes that translates into a physical $r_h$ of $11.4 \pm 1.0 $ pc. The measured half-light radius is larger than that of the discovery paper (\citealt{laevens15}; $\sim 0.4'$). To investigate this discrepancy, the sample is split between bright ($15.0 < g_0 < 23.5$) and faint ($24.0 < g_0 < 25.0$) stars, and the structural properties of Lae~3 are derived in both cases. A significant difference arises in terms of half-light radius as shown in Figure \ref{pdfs_rh}: the sample of bright stars yields a more compact size than with the faint-end of the population. Such a discrepancy would naturally arise in a satellite in which a mass-segregation process has already occurred, and could explain the difference between this work and \citet{laevens15} who analysed the system with the shallower PS1 data. To test this, the structural analysis is performed using directly the PS1 data. The resulting PDF is shown as the dashed line in Figure \ref{pdfs_rh}. The half-light radius inferred with this procedure is similar to the one obtained by L15, suggesting that the larger size derived from the MegaCam data is driven by less massive stars below $g < 22.5$ mag and that Lae~3 is mass-segregated. We compute the relaxation time of Lae~3 using the equations of \citet{koposov07} and references therein to confirm that the satellite had enough time to mass-segregate. We choose a mass-to-light ratio of 2 expected from old GCs \citep{bell_dejong01}, a total luminosity of $1125 \lsun$ determined below, and an average star mass of $0.6 \msun$. The resulting half-light relaxation time is around 2.2 Gyr, largely smaller than our inference of the age of the satellite ($13.0 \pm 1.0$ Gyr).

Two favoured stellar populations are presented in Figure \ref{cornerplots}: with and without using the priors on the metallicity and distance modulus coming respectively from the spectroscopic analysis of section 4 and the two RR Lyrae in the system. Without those priors, Lae~3 is found to be old ($13.0\pm 1.0$ Gyr) and metal-poor ($<\FeH_\mathrm{CMD}> = -2.0 \pm 0.1$ dex). The abundance ratio in $\alpha$ elements is [$\alpha$/Fe]$ = 0.2 \pm 0.2$ dex, while the distance modulus is $m-M = 19.05^{+0.02}_{-0.10}$ mag, i.e. a physical distance of $64.4^{+0.6}_{-3.0}$ kpc. This model is represented as a dashed light green line in Figure \ref{one} and nicely follows the sequence of the satellite and the spectroscopic members identified in the next section. The favoured model, i.e. the one based on the metallicity and distance priors, is similar. The structural properties, age, metallicity and $\alpha$ abundance ratio are compatible. However, the satellite is found to be closer ($m - M = 18.94^{+0.05}_{-0.02}$ mag, which translates in a physical distance of $61.4^{+1.2}_{-1.0}$ kpc) in this case. This population, represented as a solid green line in Figure \ref{one}, also follows the features of Lae~3 in the CMD. The two isochrones are barely distinguishable and the last model is the one used in the rest of this work since it is based on a spectroscopic measurement of the metallicity of the system. Using the favoured model, two quantities are defined: a ``CMD probability membership'' that assigns a probability to a given star solely based on its compatibility with the favoured stellar population of Lae~3 and a ``CMD and spatial probability membership'' that also takes the spatial location of a given star into account.

Using this CMD membership probability, we search for potential tidal structures. To do so, the field of view is spatially binned with 0.2 arcminutes bins. The CMD probability of all stars falling in a given bin are then added. This procedure therefore assigns higher values to bins that contain stars compatible with the stellar population of Lae~3. The result is shown in Figure \ref{density_plot}. This analysis shows that the satellite is highly spherical and that there is no tidal feature in the field of view compatible with the CMD properties of Lae~3.

The luminosity of the satellite is estimated following the method detailed in \citet{martin16_dra} which consists in simulating thousands of CMDs with the stellar and structural properties of Lae~3 derived earlier, and compute their resulting luminosities. This procedure yields a luminosity of $L_V = 1125^{+221}_{-129} \lsun$, translating into an absolute magnitude of $M_V = -2.8^{+0.2}_{-0.3}$ mag. This result is roughly one magnitude fainter than that found by \citet{laevens15} in the discovery paper of Lae~3. We observed a similar trend for another faint satellite discovered by \citet{laevens15}: Draco II (Dra~II). In L18, the inferred luminosity was significantly lower than found in the 2015 paper, and we concluded that it is most likely due to the overestimation of the number of giants in the system, probably due to the shallowness of the PS1 data used for the discovery of both Lae~3 and Dra~II. Though Lae~3 is clearly brighter than Dra~II, it is also significantly more distant, and the same overestimation effect might have affected the result of \citet{laevens15}, as using the same technique for a brighter MW satellite \citep{longeard19} did not yield such an effect.

\begin{table*}
\renewcommand{\arraystretch}{1.2}
\begin{center}
\caption{Inferred properties of Lae~3.\label{tbl-2}}
\begin{tabular}{lcccc}
\hline
Parameter &  Unit & Prior & Favoured model & Uncertainties  \\
\hline

RA $ \alpha $ & degrees & --- & $316.72938021$ & $\pm 0.00076375$ \\
 &  &  & 21:06:55:05 &  \\
DEC $ \delta $ & degrees & --- & $+14.98439985$  &  $\pm 0.00077118$ \\
 &  &  & +14:59:03:84 &  \\
$l$ & degrees & --- & $63.598$ & $\pm 0.001$ \\
$b$  & degrees & --- & $-21.176$  &  $\pm 0.001$ \\
$r_{h}$ & arcmin & $> 0$ & $0.64$ & $\pm 0.05$ \\
$r_{h}$ & pc &$ > 0$ & $11.4$ & $\pm 1.0$ \\
$\theta$ & degrees & [0,180] & $72$ & $^{+24}_{-17}$ \\
$\epsilon$ & --- & $> 0$ & $0.11$ & $^{+0.09}_{-0.11}$ \\
Distance modulus & mag & $G(18.88,0.04)$ & 18.94 & $^{+0.05}_{-0.02}$  \\
Distance & kpc & & $61.4$ & $^{+1.2}_{-1.0}$ \\
Age & Gyr & [8.0,13.5] & $13.0$ & $\pm 1.0$  \\
$\FeH_\mathrm{spectro}$ & dex & --- & $-1.8$ & $\pm 0.1$  \\
$[\alpha/\textrm{Fe}]$ & dex & [-0.2,0.6] & 0.0 & $\pm 0.2$ \\
M$_V$ & mag & --- & $-2.8$ & $^{+0.2}_{-0.3}$ \\
$\mu_{0}$ & mag arcsec$^{-2}$ & --- & $25.0$ & $\pm 0.3$  \\
$<v_r>$ & $\kms$ & --- & $-70.2$ & $\pm 0.5$  \\
$\mu_{\alpha}^{*}$ & mas yr$^{-1}$  & --- & $0.51$ & $\pm 0.28$   \\
$\mu_{\delta}$ & mas.yr$^{-1}$ &  --- & $-0.83$ & $\pm 0.27$   \\
Apocenter & kpc  & --- & $85.6$ & $^{+17.2}_{-5.9}$   \\
Pericenter & kpc &  --- & $40.7$ & $^{+5.6}_{-14.7}$   \\
$e_{\mathrm{orbit}}$& ---  & $> 0$ & 0.60 & $^{+0.04}_{-0.06}$  \\
U & $\kms$ & ---  & 13.1 & $^{+64.2}_{-56.4}$   \\
V & $\kms$ & ---  & $-187.3$ & $^{+45.1}_{-28.4}$  \\
W & $\kms$ & ---  & $-211.8$ & $^{+59.0}_{-46.0}$  \\
L$_z$ & km s$^{-1}$ kpc  & ---  & 793 & $^{+4010}_{-3442} $ \\
E & km$^{2}$ s$^{-2}$ & ---  & 20819& $^{+14822}_{-9163}$  \\

\hline
\end{tabular}
\end{center}
\end{table*}

\begin{figure}
\begin{center}
\centerline{\includegraphics[width=\hsize]{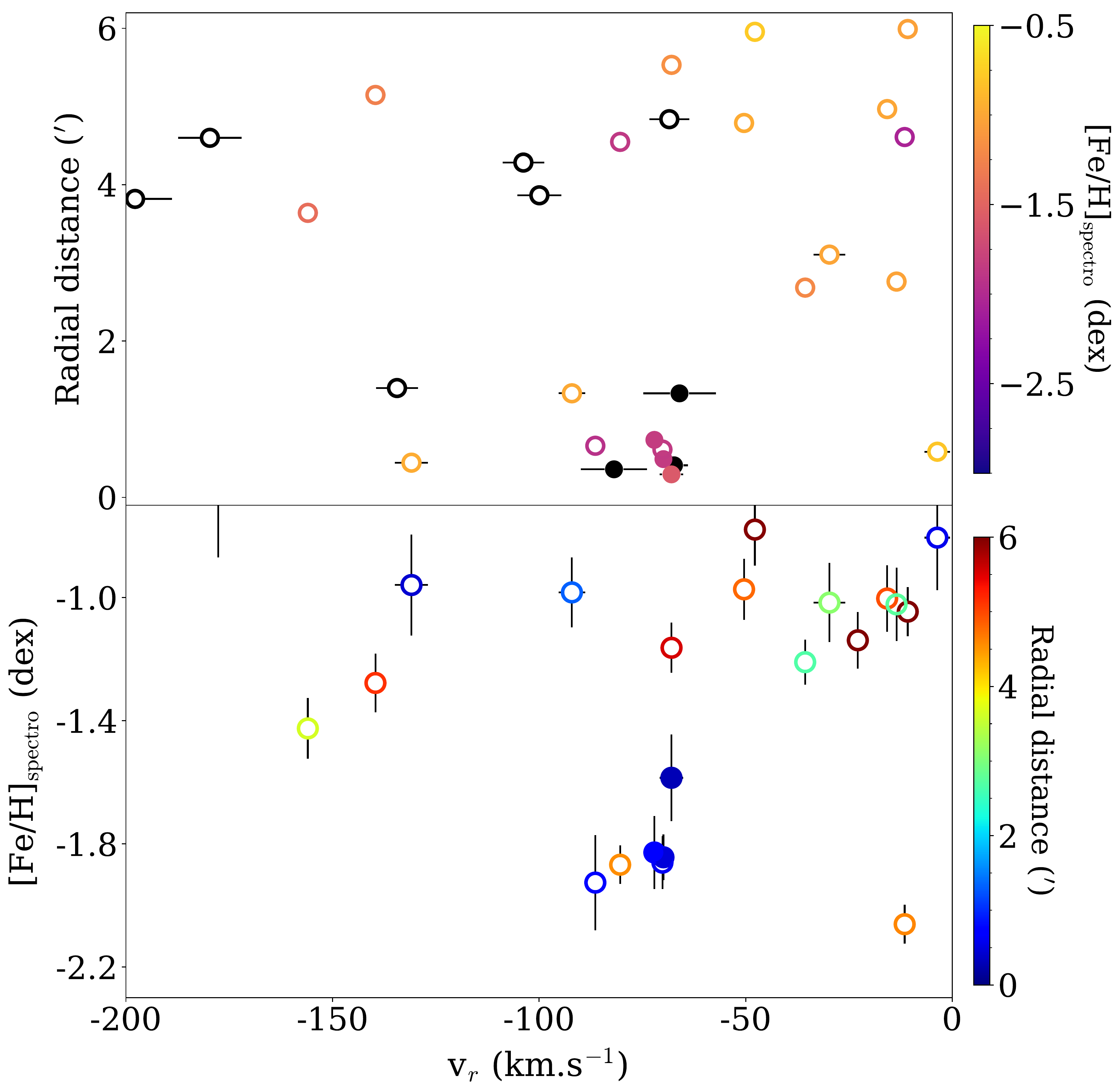}}
\caption{ Heliocentric velocities versus radial distances (top panel) and spectroscopic metallicities (bottom panel). Coloured circles are non-HB stars with a S/N greater than 10, for which we are able to derive the spectroscopic metallicities. The colormaps stand for the metallicity (top) and radial distance (bottom). The spectroscopic members are shown as filled dots.}
\label{histos_vel}
\end{center}
\end{figure}

\section{Spectroscopic analysis}

The distribution of the heliocentric velocities for all stars in our spectroscopic sample is shown in the top panel of Figure \ref{histos_vel}, along with their radial distances and spectroscopic metallicities (if possible).

\subsection{Dynamical properties}

\begin{figure*}
\begin{center}
\centerline{\includegraphics[width=\hsize]{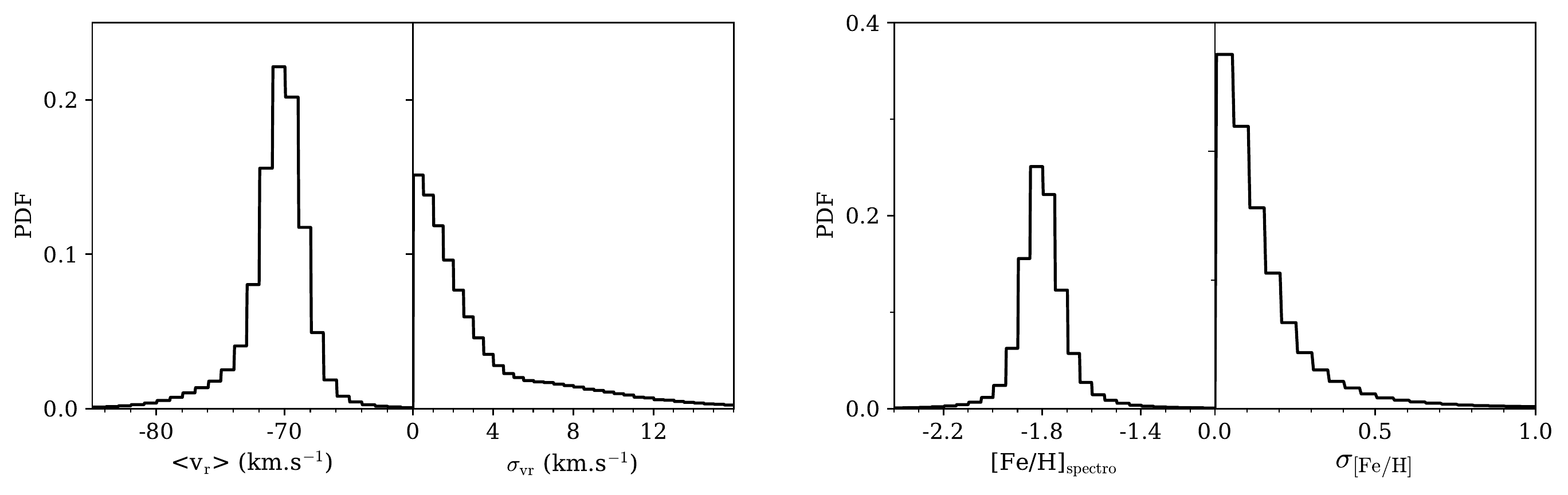}}
\caption{\textit{Left panels: }1-D marginalised PDFs of the systemic velocity and its associated dispersion. \textit{Right panels: }1-D marginalised PDFs of the systemic metallicity and its associated dispersion. The two measurements of the dispersions are unresolved.}
\label{pdfs}
\end{center}
\end{figure*}

The Lae~3 population is not prominent, and its systemic velocity overlaps that of the foreground MW stars (Figure \ref{histos_vel}). Our approach is similar to L18: the velocity distribution is assumed to be the sum of the contamination (halo and disc stars) and the Lae~3 population, both modelled with different normal distributions. To highlight Lae~3's population in the spectroscopic dataset,  the individual likelihood of each star is weighted by its spatial and CMD probability estimated from the favoured structural model of section 3 \citep{collins10}. This analysis yields a systemic radial velocity of $<$~$v_r> = -70.2 \pm 0.5 \kms$. The 1-D marginalised PDFs of the velocity parameters are represented in the left panels of Figure \ref{pdfs}. As a consequence to the low number of Lae~3 stars, the velocity dispersion is unresolved. Finally, six stars with a dynamical, structural and CMD membership probability greater than 90 per cent are identified as Lae~3 members and shown as filled circles in Figure \ref{histos_vel}.

\subsection{Spectroscopic metallicity}
 The individual metallicities of stars observed with spectroscopy can be estimated using the calibration of the Calcium triplet \citep{starkenburg10} for RGB stars, and shown in Figure \ref{histos_vel}. Member stars fainter than 21 in the $g$ band, and with S/N $<$ 10 are further discarded from our spectroscopic catalog. Only three stars are left to infer the systemic metallicity and metallicity dispersion of Lae~3, by assuming that the metallicities are normally distributed. This yields a spectroscopic metallicity of $\FeH_\mathrm{spectro} = -1.8 \pm 0.1$ dex. The same analysis is also performed using the calibration of \citet{carrera13} for metal-poor stars on the RGB and sub-RGB branch, and yields compatible results. Once more, low number statistics has a direct consequence on our ability to constrain efficiently the metallicity dispersion, which is found to be unresolved, with $\sigma_{\FeH} < 0.5$ dex at the 95 per cent confidence level.  The PDFs of both parameters are shown in the right panels of Figure \ref{pdfs}.

 \section{Gaia DR2 proper motions and orbit}
 
 \begin{figure}
\begin{center}
\centerline{\includegraphics[width=\hsize]{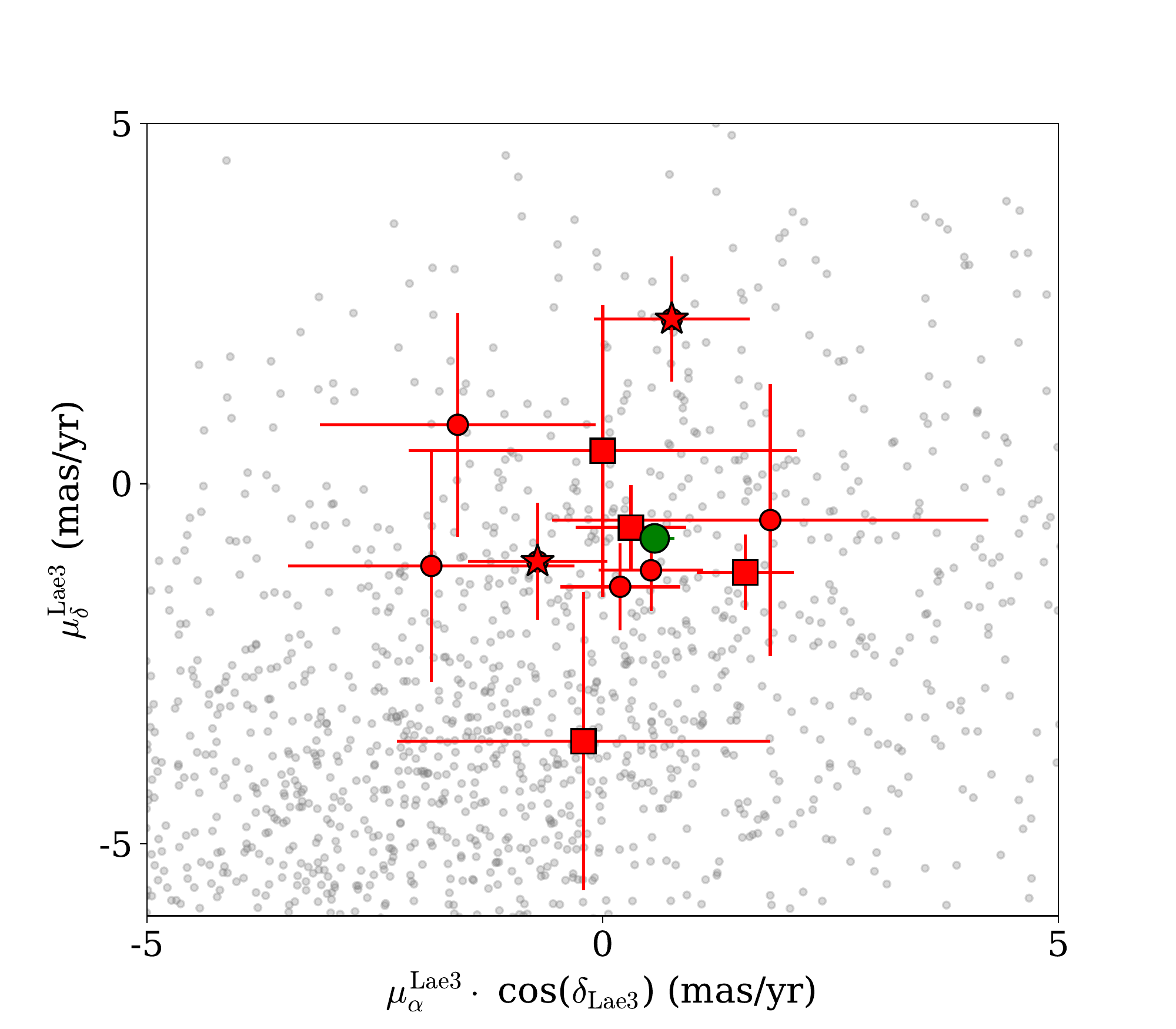}}
\caption{PMs of all stars within 15$'$ of Lae~3. The grey transparent dots show the PMs of field stars. The measurements of the four spectroscopic members with PM in Gaia DR2 are represented as squares, while the red stars and dots respectively show the PMs of the RR Lyrae stars as well as the spatially and CMD selected stars. The large green dot marks the combined PM measurement of Lae~3.}
\label{pms}
\end{center}
\end{figure}

\begin{figure*}
\begin{center}
\centerline{\includegraphics[width=\hsize]{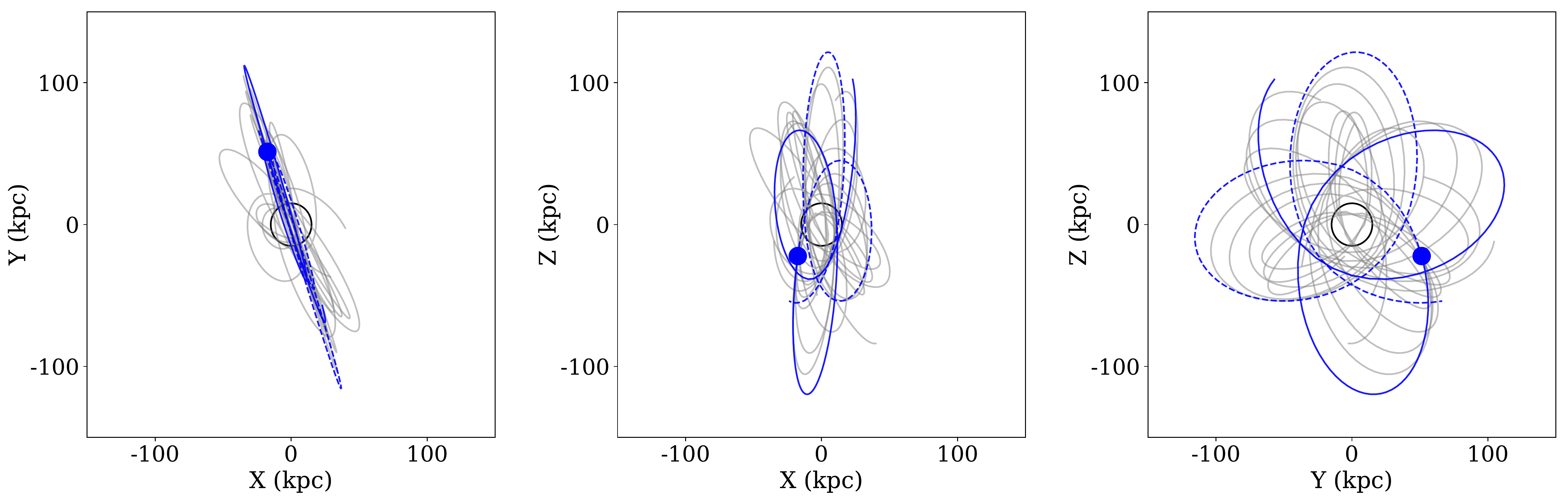}}
\caption{Orbits of Lae~3 in the X-Y, X-Z and Y-Z planes integrated over 5 Gyr. The blue line is the orbit for the favoured distance, radial velocity, position and proper motion. Grey, transparent lines are random realizations of the orbit. The MW is represented by the black circle ($R_\mathrm{MW} = 15 \kpc$), while the blue dot indicates the location of Lae~3 at present day.}
\label{orbit}
\end{center}
\end{figure*}

To infer the orbital properties of Lae~3, we cross-match all spectroscopic members and RR Lyrae stars with the Gaia Data Release 2 \citep{brown18}. Among those, four stars have a proper motion (PM) measurement in Gaia. Furthermore, all stars in the Gaia catalog with a CMD and structural membership probability greater than 90 per cent are included. Six additional stars are retrieved through this procedure, and their PMs are compatible with those of the spectroscopic members, as shown in Figure \ref{pms}. The uncertainty-weighted average PM of Lae~3 yields $\mu_{\alpha}^\mathrm{*,Lae3} = \mu_{\alpha}^\mathrm{Lae3} \cos(\delta) = 0.51 \pm 0.28$ mas\,yr$^{-1}$ and $\mu_{\delta}^\mathrm{Lae3} = -0.83 \pm 0.27$ mas\,yr$^{-1}$. These measurements take into account the systematic error of 0.035 mas.yr$^{-1}$  on the PMs for dSph as shown by \citet{helmi18}. We point out that this choice of systematic error does not change our results, given the measured uncertainties on the PM of the satellite.

We use the GALPY package \citep{bovy15} to integrate the orbit of Lae~3. The MW potential chosen to integrate Lae~3 orbit is a variant of the ``MWPotential14'' defined within GALPY, but updated with a halo mass of $1.2 \times 10^{12} \msun$ \citep{bland16}. Five thousand orbits are integrated backwards and forwards over 5 Gyr, each time by randomly drawing a position, distance, radial velocity, and PMs from their corresponding PDFs. Around 20 per cent of the resulting orbits are not bound to the MW. In the case where Lae~3 is bound to the MW, the pericenter is at $40.7 ^{+5.6}_{-14.7}$ kpc and the apocenter is at $85.6^{+17.2}_{-5.9}$ kpc. The favoured orbit of the satellite is shown as a solid blue line in Figure \ref{orbit} and corresponds to a typical outer halo orbit. In the unbound case, the apocenter is undefined and the pericenter is larger, at $59.1^{+0.7}_{-2.1}$ kpc.

\section{Discussion and conclusion}

\begin{figure*}
\begin{center}
\centerline{\includegraphics[width=\hsize]{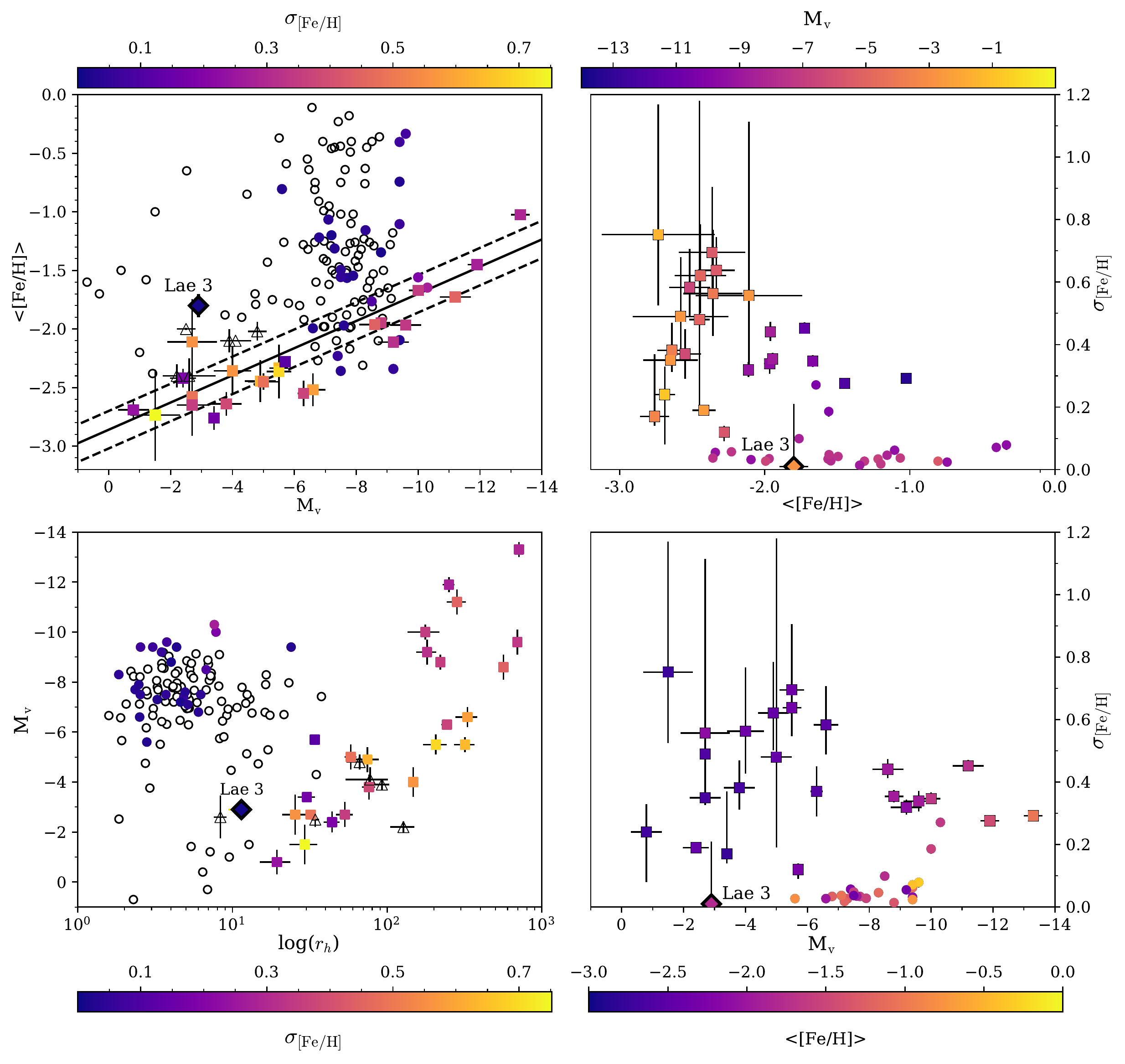}}
\caption{Comparison of Lae~3 with other GCs and dwarf galaxies of the Milky Way. Squares represent dwarf galaxies while circles represent globular clusters, and the diamond corresponds to Lae~3. Triangles stand for recently discovered dwarf-galaxy candidates that await confirmation. Hollow markers correspond to systems for which no metallicity dispersion measurement can be found in the literature. The solid line in the top-left panel corresponds to the luminosity-metallicity relation of \citet{kirby13} for dwarf spheroidals and dwarf irregulars. Dashed lines represent the RMS about this relation, also taken from \citet{kirby13}. Among the 123 globular clusters presented here, the properties of 116  were extracted from \citet{harris96} catalog, revised in 2010. For the remaining ones (Kim 1, Kim 2, Kim 3, Laevens 1, Balbinot 1, Munoz 1 and SMASH 1) parameters of the discovery publications were used (\citet{kim15b}, \citet{kim15}, \citet{kim16b}, \citet{laevens14}, \citet{balbinot13}, \citet{munoz12b} and \citet{martin16c}). Globular cluster metallicity spread measurements are taken from \citet{willman_strader12} and references therein: \citet{carretta06,carretta07b,carretta09b,carretta11}, \citet{cohen10}, \citet{gratton07}, \citet{johnson_pilachowski10}, and \citet{marino11}. \citet{mcconnachie12} and \citet{willman_strader12} are used to compile the properties of the dwarf galaxies represented here. The 18 dwarf galaxies represented here are: Bootes I \citep{belokurov06,norris10}, Canes Venatici I \citep{zucker06b}, Canes Venatici II \citep{sakamoto06}, Coma Berinices, Hercules, Leo IV and Segue I \citep{belokurov07}, Draco and Ursa Minor \citep{wilson55}, Fornax \citep{shapley38b}, Leo I and Leo II \citep{harrington_wilson50}, Pisces II \citep{belokurov10}, Sculptor \citep{shapley38a}, Sextans \citep{irwin90}, Ursa Major I \citep{willman05b}, Ursa Major II \citep{zucker06a}, Willman I \citep{willman05a}. Their metallicity and metallicity spreads were drawn from \citet{kirby08}, \citet{kirby10}, \citet{norris10}, \citet{willman11}. The dwarf galaxy candidates discovered recently and shown on this figure are Bootes II \citep{koch_rich14}, DES1 \citep{luque16,conn18}, Eridanus III \citep{bechtol15,conn18,koposov15}, Hyades II \citep{martin15}, Pegasus III \citep{kim15b}, Reticulum II and Horologium I \citep{koposov15b}, Segue II \citep{belokurov09}, and the most significant candidates of \citet{drlica-wagner15} : Gru II, Tuc III, and Tuc IV.}
\label{context}
\end{center}
\end{figure*}

We present in this paper an analysis of the faint satellite Lae~3 using deep MegaCam/CFHT broadband $g$- and $i$-band photometry of Lae~3 as well as multi-object spectroscopy observed with Keck II/DEIMOS. Lae~3 has a systemic velocity that overlaps with the MW foreground contamination: $\langle v_r \rangle = -70.2 \pm 0.5 \kms$, but an unresolved velocity dispersion. Using these results, six stars are unambiguously identified as Lae~3 members, and three are bright enough to be used to estimate the systemic metallicity of the satellite. Lae~3 comes out as a fairly metal-poor stellar system: $\langle\FeH_\mathrm{spectro}\rangle = -1.8 \pm 0.1$ dex which places Lae~3 far off the luminosity-metallicity relation of DGs \citep{kirby13} as shown in Figure \ref{context}. The metallicity dispersion is also unresolved. Similarly to \citet{laevens15}, two RR Lyrae stars are used to estimate the distance of Lae~3, and yield a distance modulus of $18.88 \pm 0.04$ mag. Using these results as priors, we derive the structural and CMD properties and find a half-light radius of $11.4 \pm 1.0$ pc, a marginally resolved ellipticity and a final distance modulus measurement of $18.94^{+0.05}_{-0.02}$ mag. A discrepancy between the half-light radius of Lae~3 derived using bright and faint stars hints that the satellite is mass-segregated. This hypothesis is strengthened by the relaxation time of the satellite of $\sim 2.2$ Gyr, much smaller than the age of the satellite found to be $13.0 \pm 1.0$ Gyr by our CMD fitting procedure. The sphericity of Lae~3 and an analysis of the density of Lae~3-like stars in the field show no clear sign of tidal features that might hint at a perturbation of the system and therefore its ability to mass-segregate. The favoured stellar population is metal-poor, not particularly enriched in $\alpha$ elements, and at a distance of $61.4^{+1.2}_{-1.0}$ kpc. Finally, the orbit calculation yields an outer halo orbit, with a pericenter of $40.7 ^{+5.6}_{-14.7}$ kpc and an apocenter of $85.6 ^{+17.2}_{-5.9}$ kpc.

Lae~3 shows the main characteristics of MW globular clusters: the satellite is fairly spherical and is at the same time more compact and metal-rich than DGs of the same luminosity \citep{mcconnachie12, kirby13}, such as Ret~II ($M_V \sim -2.7$), Hor~I ($M_V \sim -3.4$) or Boo~II ($M_V \sim -2.7$) as shown in the bottom-left panel of Figure \ref{context}. Regarding the size and galactocentric distance of the satellite, Lae~3 can be compared to SMASH~1 \citep{martin16c}. SMASH 1 has a size of $9.1^{+5.9}_{-3.4}$ pc, and is lying at $\sim 57 \kpc$ of the center of the galaxy. The location and distance of SMASH~1 imply that it may be a satellite of the LMC. However, Lae~3 is brighter ($-2.8$ vs $-1.0$ mag) and is more metal-rich ($-1.8$ vs $-2.2$ dex). The top-left panel of Figure \ref{context} shows that the systemic metallicity of Lae~3 is offset by $\sim 0.7$ dex from the metallicity-luminosity relation of dwarf galaxies \citep{kirby13}. We have to turn to Pal 1 or Pal 13 \citep{harris10} to find a cluster with a luminosity comparable to the one of Lae~3 (respectively of $\sim -2.5$ and $\sim -3.8$ mag). Still, these two GCs are much more compact, with a size of the order of the parsec. 

Both the velocity and metallicity dispersions of Lae~3 are unresolved, although the small number of member stars in our spectroscopic dataset does not give stringent enough constraints to rule out a dynamically hot system or that it is chemically enriched (right panels of Figure \ref{context}). Lae~3 is possibly mass-segregated, which implies that its internal dynamics is ruled by purely baryonic two-bodies interactions \citep{kim15} and it is statistically incompatible with the luminosity-metallicity relation of DGs. We therefore conclude that Lae~3 likely is a MW outer halo globular cluster.

\section{Acknowledgments}
NL, NFM, and RI gratefully acknowledge support from the French National Research Agency (ANR) funded project ``Pristine'' (ANR-18-CE31-0017) along with funding from CNRS/INSU through the Programme National Galaxies et Cosmologie and through the CNRS grant PICS07708. ES, KY, NM, AA, JIGH, and NL benefited from the International Space Science Institute (ISSI) in Bern, CH, thanks to the funding of the Teams ``The Formation and Evolution of the Galactic Halo'' and ``Pristine''. This work has been published under the framework of the IdEx Unistra and benefits from a funding from the state managed by the French National Research Agency as part of the investments for the future program. This research was supported in part by the National Science Foundation under Grant No. NSF PHY11-25915. DM is supported by an Australian Research Council (ARC) Future Fellowship (FT160100206). BPML gratefully acknowledges support from FONDECYT postdoctoral fellowship No. 3160510.

We gratefully thank the CFHT staff for performing the observations in queue mode, for their reactivity in adapting the schedule, and for answering our questions during the data-reduction process. We thank Nina Hernitschek for granting us access to the catalogue of Pan-STARRS variability catalogue. 

Based on observations obtained at the Canada-France-Hawaii Telescope (CFHT) which is operated by the National Research Council of Canada, the Institut National des Sciences de l'Univers of the Centre National de la Recherche Scientifique of France, and the University of Hawaii.

Some of the data presented herein were obtained at the W. M. Keck Observatory, which is operated as a scientific partnership among the California Institute of Technology, the University of California and the National Aeronautics and Space Administration. The Observatory was made possible by the generous financial support of the W. M. Keck Foundation. Furthermore, the authors wish to recognize and acknowledge the very significant cultural role and reverence that the summit of Maunakea has always had within the indigenous Hawaiian community.  We are most fortunate to have the opportunity to conduct observations from this mountain.

The Pan-STARRS1 Surveys (PS1) have been made possible through contributions of the Institute for Astronomy, the University of Hawaii, the Pan-STARRS Project Office, the Max-Planck Society and its participating institutes, the Max Planck Institute for Astronomy, Heidelberg and the Max Planck Institute for Extraterrestrial Physics, Garching, The Johns Hopkins University, Durham University, the University of Edinburgh, Queen's University Belfast, the Harvard-Smithsonian Center for Astrophysics, the Las Cumbres Observatory Global Telescope Network Incorporated, the National Central University of Taiwan, the Space Telescope Science Institute, the National Aeronautics and Space Administration under Grant No. NNX08AR22G issued through the Planetary Science Division of the NASA Science Mission Directorate, the National Science Foundation under Grant No. AST-1238877, the University of Maryland, and Eotvos Lorand University (ELTE).

This work has made use of data from the European Space Agency (ESA)
mission {\it Gaia} (\url{https://www.cosmos.esa.int/gaia}), processed by
the {\it Gaia} Data Processing and Analysis Consortium (DPAC,
\url{https://www.cosmos.esa.int/web/gaia/dpac/consortium}). Funding
for the DPAC has been provided by national institutions, in particular
the institutions participating in the {\it Gaia} Multilateral Agreement.

\newpage

\begin{table*}
\caption{Properties of our spectroscopic sample. Confirmed members are denoted by ``Y'' and non-members by ``N''. The star denoted ``Y?'' is a plausible member, as its position, spectroscopic metallicity and velocity are compatible with those of Lae~3. However, it is not confirmed as it does not pass our membership probability cut ($P_{mem} \geq 90$ \%).
\label{tbl-2}}

\setlength{\tabcolsep}{4.5pt}
\renewcommand{\arraystretch}{0.4}
\begin{sideways}
\begin{tabular}{cccccccccccccc}
\hline
RA (deg) & DEC (deg) & $g_0$ & $i_0$ & $v_{r} (\kms)$ & $\mu_{\alpha}^{*}$ (mas.yr$^{-1}$) & $\mu_{\delta}$ (mas.yr$^{-1}$) &  S/N & [Fe/H]$_\mathrm{spectro}$ & $P_\mathrm{mem}$ & Member\\
\hline

316.71305417 & 15.01271111 & 17.94 $\pm$ 0.01 & 17.23 $\pm$ 0.01 & -35.6 $\pm$ 1.4 & 3.647 $\pm$ 0.221 & -3.548 $\pm$ 0.242 & 39.3 & -1.21 $\pm$ 0.07 & 0.00 & N \\ \\ 
316.68967917 & 15.06302222 & 18.19 $\pm$ 0.01 & 17.41 $\pm$ 0.01 & -10.8 $\pm$ 1.4 & -0.763 $\pm$ 0.239 & -0.821 $\pm$ 0.254 & 36.3 & -1.05 $\pm$ 0.08 & 0.00 & N \\ \\ 
316.67275417 & 15.03253056 & 18.73 $\pm$ 0.01 & 18.13 $\pm$ 0.01 & -15.8 $\pm$ 1.7 & -5.232 $\pm$ 0.413 & -8.422 $\pm$ 0.403 & 22.1 & -1.0 $\pm$ 0.11 & 0.00 & N \\ \\ 
316.68561250 & 15.01203889 & 18.96 $\pm$ 0.01 & 18.34 $\pm$ 0.01 & 8.1 $\pm$ 2.1 & -2.672 $\pm$ 0.44 & -6.607 $\pm$ 0.465 & 21.9 & -1.15 $\pm$ 0.1 & 0.00 & N \\ \\ 
316.68863750 & 15.05416944 & 19.01 $\pm$ 0.01 & 18.47 $\pm$ 0.01 & -68.0 $\pm$ 1.4 & -1.407 $\pm$ 0.459 & -2.176 $\pm$ 0.464 & 26.9 & -1.16 $\pm$ 0.08 & 0.00 & N \\ \\ 
316.69020000 & 15.01787500 & 19.03 $\pm$ 0.01 & 18.60 $\pm$ 0.01 & -156.0 $\pm$ 1.6 & -2.898 $\pm$ 0.505 & -7.339 $\pm$ 0.511 & 20.0 & -1.43 $\pm$ 0.1 & 0.00 & N \\ \\ 
316.68378750 & 15.04429167 & 19.48 $\pm$ 0.01 & 18.89 $\pm$ 0.01 & -139.6 $\pm$ 1.6 & -2.633 $\pm$ 0.625 & -13.336 $\pm$ 0.613 & 23.3 & -1.28 $\pm$ 0.09 & 0.00 & N \\ \\ 
316.71363750 & 15.01609167 & 19.53 $\pm$ 0.01 & 19.41 $\pm$ 0.01 & 429.0 $\pm$ 2.6 & -0.284 $\pm$ 0.873 & -0.101 $\pm$ 0.927 & 20.8 & -2.95 $\pm$ 0.14 & 0.00 & N \\ \\ 
316.75032500 & 15.01542778 & 19.59 $\pm$ 0.01 & 19.23 $\pm$ 0.01 & -346.7 $\pm$ 1.6 & -2.236 $\pm$ 0.853 & -1.489 $\pm$ 0.947 & 24.0 & -1.96 $\pm$ 0.08 & 0.00 & N \\ \\ 
316.70143333 & 15.01480278 & 19.81 $\pm$ 0.01 & 19.15 $\pm$ 0.01 & -29.7 $\pm$ 3.8 & -4.405 $\pm$ 0.758 & -0.392 $\pm$ 0.796 & 18.5 & -1.02 $\pm$ 0.13 & 0.01 & N \\ \\ 
316.69964167 & 15.01039167 & 21.29 $\pm$ 0.01 & 20.83 $\pm$ 0.01 & -515.3 $\pm$ 3.0 & --- $\pm$ --- & --- $\pm$ --- & 9.1 & --- $\pm$ --- & 0.00 & N \\ \\ 
316.69013750 & 15.03111667 & 21.61 $\pm$ 0.01 & 21.05 $\pm$ 0.01 & -103.8 $\pm$ 5.0 & --- $\pm$ --- & --- $\pm$ --- & 8.0 & --- $\pm$ --- & 0.01 & N \\ \\ 
316.72665833 & 15.05161389 & 21.92 $\pm$ 0.01 & 21.52 $\pm$ 0.01 & -331.2 $\pm$ 3.6 & --- $\pm$ --- & --- $\pm$ --- & 5.4 & --- $\pm$ --- & 0.00 & N \\ \\ 
316.71725417 & 15.03437222 & 22.00 $\pm$ 0.01 & 21.38 $\pm$ 0.01 & -99.9 $\pm$ 5.3 & --- $\pm$ --- & --- $\pm$ --- & 6.4 & --- $\pm$ --- & 0.01 & N \\ \\ 
316.72435000 & 14.99263333 & 22.31 $\pm$ 0.02 & 21.93 $\pm$ 0.02 & -66.0 $\pm$ 8.8 & --- $\pm$ --- & --- $\pm$ --- & 4.3 & --- $\pm$ --- & 0.99 & Y \\ \\ 
316.66194583 & 15.06772222 & 18.37 $\pm$ 0.01 & 17.62 $\pm$ 0.01 & -22.9 $\pm$ 1.4 & 1.033 $\pm$ 0.276 & 0.696 $\pm$ 0.289 & 30.0 & -1.14 $\pm$ 0.09 & 0.00 & N \\ \\ 
316.66891250 & 15.02001944 & 18.50 $\pm$ 0.01 & 17.75 $\pm$ 0.01 & -11.5 $\pm$ 1.1 & 1.047 $\pm$ 0.322 & -1.302 $\pm$ 0.317 & 29.3 & -2.06 $\pm$ 0.06 & 0.00 & N \\ \\ 
316.65947083 & 15.04309444 & 19.85 $\pm$ 0.01 & 19.08 $\pm$ 0.01 & -47.8 $\pm$ 1.9 & -3.144 $\pm$ 0.76 & -4.985 $\pm$ 0.724 & 20.7 & -0.78 $\pm$ 0.12 & 0.00 & N \\ \\ 
316.66246667 & 15.12159167 & 20.13 $\pm$ 0.01 & 19.55 $\pm$ 0.01 & -177.7 $\pm$ 2.8 & 0.483 $\pm$ 1.105 & -6.195 $\pm$ 0.963 & 21.7 & -0.61 $\pm$ 0.26 & 0.00 & N \\ \\ 
316.66070833 & 15.07443611 & 21.24 $\pm$ 0.01 & 20.68 $\pm$ 0.01 & -138.4 $\pm$ 4.7 & --- $\pm$ --- & --- $\pm$ --- & 9.5 & --- $\pm$ --- & 0.00 & N \\ \\ 
316.63902917 & 15.05994444 & 21.56 $\pm$ 0.01 & 21.12 $\pm$ 0.01 & -274.5 $\pm$ 11.8 & --- $\pm$ --- & --- $\pm$ --- & 5.9 & --- $\pm$ --- & 0.00 & N \\ \\ 
316.65706250 & 15.06192222 & 22.07 $\pm$ 0.01 & 21.49 $\pm$ 0.01 & -131.8 $\pm$ 6.9 & --- $\pm$ --- & --- $\pm$ --- & 5.0 & --- $\pm$ --- & 0.00 & N \\ \\ 
316.80862500 & 14.94329722 & 21.05 $\pm$ 0.01 & 20.35 $\pm$ 0.01 & -68.5 $\pm$ 4.8 & --- $\pm$ --- & --- $\pm$ --- & 12.2 & --- $\pm$ --- & 0.00 & N \\ \\ 
316.79680833 & 14.89963056 & 21.38 $\pm$ 0.01 & 20.71 $\pm$ 0.01 & -221.0 $\pm$ 4.5 & --- $\pm$ --- & --- $\pm$ --- & 8.7 & --- $\pm$ --- & 0.00 & N \\ \\ 
316.79215417 & 14.94927500 & 22.47 $\pm$ 0.02 & 22.11 $\pm$ 0.02 & -197.8 $\pm$ 8.9 & --- $\pm$ --- & --- $\pm$ --- & 3.3 & --- $\pm$ --- & 0.13 & N \\ \\ 
316.75973333 & 14.90096111 & 19.07 $\pm$ 0.01 & 18.68 $\pm$ 0.01 & -80.4 $\pm$ 2.0 & 0.335 $\pm$ 0.641 & -2.754 $\pm$ 0.557 & 26.2 & -1.87 $\pm$ 0.06 & 0.00 & N \\ \\ 
316.77516250 & 14.90417500 & 19.22 $\pm$ 0.01 & 18.55 $\pm$ 0.01 & -50.4 $\pm$ 1.9 & -1.68 $\pm$ 0.596 & -8.502 $\pm$ 0.498 & 27.5 & -0.97 $\pm$ 0.1 & 0.00 & N \\ \\ 
316.72701667 & 14.97872778 & 19.24 $\pm$ 0.01 & 18.49 $\pm$ 0.01 & -69.9 $\pm$ 1.6 & 1.565 $\pm$ 0.53 & -1.232 $\pm$ 0.523 & 23.3 & -1.84 $\pm$ 0.07 & 0.98 & Y \\ \\ 
316.72812500 & 14.98119167 & 19.31 $\pm$ 0.01 & 18.63 $\pm$ 0.01 & -70.2 $\pm$ 1.4 & 0.31 $\pm$ 0.608 & -0.609 $\pm$ 0.592 & 22.5 & -1.86 $\pm$ 0.09 & 0.36 & Y? \\ \\ 
316.72099583 & 14.97752500 & 19.58 $\pm$ 0.01 & 19.39 $\pm$ 0.01 & -86.4 $\pm$ 2.1 & 0.758 $\pm$ 0.852 & 2.285 $\pm$ 0.871 & 20.0 & -1.93 $\pm$ 0.15 & 0.00 & N \\ \\ 
316.76855417 & 14.94703889 & 19.62 $\pm$ 0.01 & 18.90 $\pm$ 0.01 & 2.9 $\pm$ 1.8 & -4.84 $\pm$ 0.79 & -3.869 $\pm$ 0.784 & 25.6 & -0.83 $\pm$ 0.1 & 0.16 & N \\ \\ 
316.76643750 & 14.94122778 & 20.09 $\pm$ 0.01 & 19.45 $\pm$ 0.01 & -13.5 $\pm$ 2.3 & -2.702 $\pm$ 1.02 & -5.581 $\pm$ 1.006 & 19.0 & -1.02 $\pm$ 0.12 & 0.02 & N \\ \\ 
316.73886667 & 14.96605556 & 20.57 $\pm$ 0.01 & 19.84 $\pm$ 0.01 & -3.6 $\pm$ 3.1 & -1.88 $\pm$ 1.572 & -1.143 $\pm$ 1.61 & 15.6 & -0.81 $\pm$ 0.17 & 0.45 & N \\ \\ 
316.72408750 & 14.96693889 & 20.59 $\pm$ 0.01 & 20.02 $\pm$ 0.01 & -130.9 $\pm$ 4.0 & 1.839 $\pm$ 2.395 & -0.506 $\pm$ 1.892 & 14.5 & -0.96 $\pm$ 0.16 & 0.63 & N \\ \\ 
316.72444583 & 14.98208333 & 20.61 $\pm$ 0.01 & 20.00 $\pm$ 0.01 & -72.1 $\pm$ 1.7 & -0.0 $\pm$ 2.13 & 0.456 $\pm$ 2.024 & 15.0 & -1.83 $\pm$ 0.12 & 0.99 & Y \\ \\ 
316.73362500 & 14.94926389 & 20.72 $\pm$ 0.01 & 20.04 $\pm$ 0.01 & -92.1 $\pm$ 3.2 & -0.211 $\pm$ 2.05 & -3.577 $\pm$ 2.068 & 14.5 & -0.98 $\pm$ 0.11 & 0.86 & N \\ \\ 
316.73480000 & 14.96889167 & 20.93 $\pm$ 0.01 & 20.29 $\pm$ 0.01 & -68.0 $\pm$ 2.9 & --- $\pm$ --- & --- $\pm$ --- & 13.3 & -1.59 $\pm$ 0.14 & 1.00 & Y \\ \\ 
316.74872917 & 14.95614167 & 21.61 $\pm$ 0.01 & 21.11 $\pm$ 0.01 & -134.4 $\pm$ 5.1 & --- $\pm$ --- & --- $\pm$ --- & 8.2 & --- $\pm$ --- & 0.09 & N \\ \\ 
316.72737500 & 14.97743611 & 21.61 $\pm$ 0.01 & 21.04 $\pm$ 0.01 & -67.4 $\pm$ 3.4 & --- $\pm$ --- & --- $\pm$ --- & 7.5 & --- $\pm$ --- & 1.00 & Y \\ \\ 
316.78203333 & 14.91307778 & 22.17 $\pm$ 0.01 & 21.56 $\pm$ 0.01 & -179.7 $\pm$ 7.6 & --- $\pm$ --- & --- $\pm$ --- & 5.3 & --- $\pm$ --- & 0.00 & N \\ \\ 
316.73203333 & 14.97697778 & 22.30 $\pm$ 0.02 & 21.74 $\pm$ 0.02 & -81.9 $\pm$ 8.0 & --- $\pm$ --- & --- $\pm$ --- & 4.5 & --- $\pm$ --- & 1.00 & Y \\ \\ 
316.72857917 & 14.95332778 & 22.48 $\pm$ 0.02 & 22.12 $\pm$ 0.02 & 60.1 $\pm$ 9.2 & --- $\pm$ --- & --- $\pm$ --- & 3.5 & --- $\pm$ --- & 1.00 & N \\ \\ 
316.77040833 & 14.98083611 & 22.52 $\pm$ 0.02 & 22.05 $\pm$ 0.02 & -437.6 $\pm$ 3.9 & --- $\pm$ --- & --- $\pm$ --- & 4.1 & --- $\pm$ --- & 0.07 & N \\ \\ 
316.75475417 & 14.97596389 & 22.54 $\pm$ 0.02 & 22.28 $\pm$ 0.02 & 738.4 $\pm$ 15.1 & --- $\pm$ --- & --- $\pm$ --- & 3.2 & --- $\pm$ --- & 1.00 & N \\ \\

\end{tabular}
\end{sideways}
\end{table*}

\newcommand{\mnras}{MNRAS}
\newcommand{\pasa}{PASA}
\newcommand{\nat}{Nature}
\newcommand{\araa}{ARAA}
\newcommand{\aj}{AJ}
\newcommand{\apj}{ApJ}
\newcommand{\apjl}{ApJ}
\newcommand{\apjs}{ApJSupp}
\newcommand{\aap}{A\&A}
\newcommand{\aaps}{A\&ASupp}
\newcommand{\pasp}{PASP}

%\bibliography{/Users/longeard/Documents/biblio}
%\bibliographystyle{mn2e}

\clearpage

\end{document}